\documentstyle[12pt,amssymb,amsfonts]{article}
\advance \textheight by \topskip
\makeatletter
\def\eqnarray{\stepcounter{equation}\let\@currentlabel=\theequation
\global\@eqnswtrue
\global\@eqcnt\z@\tabskip\@centering\let\\=\@eqncr
$$\halign to \displaywidth\bgroup\@eqnsel\hskip\@centering
  $\displaystyle\tabskip\z@{##}$&\global\@eqcnt\@ne
  \hfil$\displaystyle{\hbox{}##\hbox{}}$\hfil
  &\global\@eqcnt\tw@ $\displaystyle\tabskip\z@
  {##}$\hfil\tabskip\@centering&\llap{##}\tabskip\z@\cr}
\@addtoreset{equation}{section}
  \def\theequation{\thesection.\arabic{equation}}
\makeatother
\def\inprod{\mathop{\kern -0.05em\raise -0.1em\hbox{%
  \vrule height 0.03em width 0.6em depth 0em%
  \vrule height 0.7em width 0.03em depth 0em}\kern 0.1em}\nolimits}
\def\d{\mbox{\sf d}}
\def\Rn{{\Bbb R}^n}
\def\be{\begin{equation}}
\def\ee{\end{equation}}
\begin{document}

\begin{titlepage}
\hbox to \hsize{\hfil gr-qc/9809058}
\hbox to \hsize{\hfil September, 1998}
\vfill
\large \bf
\begin{center}
Free Boundary Poisson Bracket Algebra in Ashtekar's Formalism
\end{center}
\vskip 1cm
\normalsize
\begin{center}
{\bf Vladimir O. Soloviev\footnote{e--mail:
soloviev@th1.ihep.su}}\\
{\small Institute for High Energy Physics,\\
142284 Protvino, Moscow Region, Russia}
\end{center}
\vskip 2.cm

\begin{abstract}
\noindent
We consider the algebra of spatial diffeomorphisms and gauge transformations in the canonical formalism of  General Relativity in the Ashtekar and ADM variables. Modifying the Poisson bracket by including surface terms in accordance with our previous proposal allows us to consider all local functionals as differentiable. We show that closure of the algebra under consideration can be achieved by choosing surface terms in the expressions for the generators prior to imposing any boundary conditions. An essential point is that the Poisson structure in the Ashtekar formalism differs from the canonical one by boundary terms. 
\end{abstract}
\vfill
\end{titlepage}

\section{Introduction} 

 In the field theory  Hamiltonian formalism it is conventional to assume that well-defined Poisson brackets exist only for the so-called ``differentiable'' functionals \cite{RT, FT} whose variation does not involve a boundary contribution. However, this restriction is not necessary, as can be seen from what follows. Within different approaches, the authors of \cite{LMMR, FT85, APP} proposed this restriction to be by-passed and  extended the Hamiltonian formalism to a larger class of functionals --- and, hence, to a larger class of problems. To realize this program  different proposals have been put forward in order to modify the standard formula for the Poisson bracket by boundary terms. The authors of  \cite{FT85, APP} employed a {\it nonlocal} formula for these terms, which indicates the existence of the interaction between the Korteweg-de Vries fields at infinitely distant points on the boundary of the one-dimensional space.  In contrast, in \cite{LMMR} only {\it local} boundary terms were considered for the description of the  ideal fluid hydrodynamics, similar to those that arise when integrating the total divergences. 

Recently \cite{Sol93, Sol95} we showed that the approach proposed in \cite{LMMR} can be generalized to include arbitrary local functionals depending on an arbitrary, but finite number of spatial derivatives. The standard expression for the Poisson bracket gets an additional contribution that consists of a sum (which is formally infinite, but actually terminates after a finite number of terms) of some divergences. It turns out that the Jacobi identity, the antisymmetry, and the closedness properties are satisfied by the new Poisson bracket {\it exactly, rather than up to boundary terms}. An interesting point is that these standard requirements can be satisfied even before one imposes any boundary conditions. Thus, the new formula determines the Poisson bracket on the set of all  local functionals, not only for the ``differentiable'' functionals, as it is in the conventional approach. Thus, we hope that our results can be applied to various physical problems. Regarding the boundary conditions, our point of view is that they can be taken into account in the next step, as is the case with the gauge conditions, and result only in a subsequent reduction of the original Poisson structure. 

In this paper we use the new definition of the Poisson bracket \cite{Sol93, Sol95}  to study the canonical formalism for General Relativity in the Arnowitt-Deser-Misner (ADM) and Ashtekar variables. We show that the well-known algebra can be realized in this framework, irrespective to the choice of boundary conditions. In the Ashtekar formalism one uses a transformation of variables that  differs from the canonical one if surface terms  are taken into account   \cite{Sol92}. The deviation from the canonical relations on the boundary was also discussed  in  \cite{Smol}. 

This paper is organized as follows. We start with a brief summary  of the Hamiltonian formalism for gravity in the ADM variables  \cite{ADM}. Then we analyze the change of variables leading to the Ashtekar formalism. Next, we give the motivation for the new formula of the field theory  Poisson bracket and illustrate applications of the new formula  in the ADM approach. The transformation to the Ashtekar variables gives rise to an unconventional surface contribution that makes the transformation non-canonical. However, it is this contribution that allows us to preserve the algebra of the generators found in the ADM variables. 

\section{ADM formalism}

Space-time can be considered as a 4-manifold arising as a result of the time evolution of a three-dimensional space-like hypersurface. The dynamical variables are then the Riemannian metric tensor field $\gamma_{ij}(x^k)$, whence $\gamma_{ij}dx^idx^j\ge 0$, and the tensor density field of the conjugate momenta $\pi^{ij}(x^k)$, which are linearly related to the extrinsic curvature tensor $K_{ij}(x^k)$ of the hypersurface, 
\be
\pi^{ij}=-\sqrt{\gamma}(K^{ij}-\gamma^{ij}K),
\ee 
where $\gamma^{ij}$ is the inverse matrix to $\gamma_{ij}$, $K=\gamma^{ij}K_{ij}$, $\gamma={\rm det}|\gamma_{ij}|$, and the Latin indices label spatial coordinates and run over the values $1, 2, 3$. The summation symbol is omitted. 

The direction of the evolution is specified at each point of the hypersurface by a time-like 4-vector $N^\alpha$, whose components $N(x^k,t)$ and $N^i(x^k,t)$ satisfy the inequality 
\be
N^2\ge \gamma_{ij} N^iN^j.
\ee 
Dynamic equations are generated by the Hamiltonian 
\be
H=\int\limits_{\Omega}\left(N{\cal H}+N^i{\cal H}_i\right)d^3x,\label{eq:ham}
\ee 
which is a linear combination of the constraints 
\be
{\cal H}=-\frac{1}{\sqrt{\gamma}}\left(\gamma R+\frac{\pi^2}{2}-
{\rm Sp}\pi^2\right),\quad {\cal H}_i=-2\pi^j_{i|j},
\ee
and by the canonical Poisson bracket 
\be
\left\{\gamma_{ij}(x),\pi^{kl}(y)\right\}=\delta^{kl}_{ij}\delta(x,y)\equiv
\frac{1}{2}\left(\delta^k_i
\delta^l_j+\delta^k_j\delta^l_i\right)\delta(x,y),
\ee
such that 
\be
\dot\gamma_{ij}=\{\gamma_{ij},H\},\quad \dot\pi^{ij}=\{\pi^{ij},H\}.\label
{eq:heq}
\ee
In order to explicitly reconstruct the four-dimensional space-time geometry in arbitrary coordinates $X^{\alpha}$, $\alpha=0,1,2,3$, one should specify four additional functions relating the coordinates to $(x^k,t)$. 
\be
X^{\alpha}=e^{\alpha}(x^k,t).\label{eq:hyper}
\ee
If $t_1$ corresponds to the initial moment and $t_2$ to the final moment of the evolution, Eqs. (\ref{eq:hyper}) determine the embedding of the fixed-$t$ hypersurface for every $t_1\le t\le t_2$ into space-time. Functions $N(x^k,t), N^i(x^k,t)$ should then be thought of as the components of the 4-vector 
$N^{\alpha}=\dot e^{\alpha}$
\be
N^{\alpha}=Nn^{\alpha}+N^ie^{\alpha}_i,\label{eq:time}
\ee 
with respect to the basis $\left(n^{\alpha}, e^{\alpha}_i\right)$ at every point of the $t={\rm const}$ hypersurface, where 
\be
e^{\alpha}_i=\frac{\partial e^{\alpha}}{\partial x^i},\quad n_{\alpha}
e^{\alpha}_i=0,\quad n^{\alpha}n_{\alpha}=-1.
\ee
Having found the components of the unit normal vector $n^{\alpha}$ from Eqs. (\ref{eq:time}), we obtain the expression for the space-time metric tensor 
\be
g^{\alpha\beta}=-n^{\alpha}n^{\beta}+e^{\alpha}_ie^{\beta}_j\gamma^{ij}.
\ee 
A more detailed discussion of this covariant Hamiltonian formalism can be found in \cite{Kuch}. 

Equations (\ref{eq:heq}) are suitable not only for the purposes of describing the time evolution, but also for describing the transformations of spatial coordinates on a fixed hypersurface when $N=0$: 
\begin{eqnarray}
\dot\gamma_{ij}&=&{\cal L}_{\vec N}\gamma_{ij}=N_{i|j}+N_{j|i},\\
\dot\pi^{ij}&=&{\cal L}_{\vec N}\pi^{ij}=-N^i_{\ |k}\pi^{kj}-N^j_{\ |k}
\pi^{ik}+{\left(N^k\pi^{ij}\right)}_{|k},
\end{eqnarray} 
where ${\cal L}_{\vec N}$ is the Lie derivative along the direction of the vector field $N^i$. Obviously, the $t$ variable cannot be considered as time in this case. Let us note that the relation 
\begin{equation}
\left\{H(N,N^i), H(M,M^j)\right\}=H(L,L^k), \label{eq:alg}
\end{equation}
where
\begin{eqnarray}
L&=&N^iM_{,i}-M^iN_{,i},\nonumber \\
L^k&=&\gamma^{ki}(NM_{,i}-MN_{,i})+N^iM^k_{\
,i}-M^iN^k_{\ ,i}\label{eq:nonalg}
\end{eqnarray}
is satisfied by the Hamiltonian considered as the generator of the coordinate transformations of $(x^k,t)$, determined by the functions $N, N^i$. In ``local'' language, this is usually represented in the form 
 \begin{eqnarray}
 \left\{{\cal
 H}(x),{\cal H}(y)\right\}&=&\left(\gamma^{ik}(x){\cal H}_k(x)+
 \gamma^{ik}(y){\cal H}_k(y)\right)\delta_{,i}(x,y),\nonumber\\
 \left\{{\cal H}_i(x),{\cal H}_k(y)\right\}&=&{\cal H}_i(y)\delta_{,k}(x,y)+
 {\cal H}_k(x)\delta_{,i}(x,y),\nonumber\\
 \left\{{\cal H}_i(x),{\cal H}(y)\right\}&=&{\cal
 H}(x)\delta_{,i}(x,y).\label{eq:algconstr}
\end{eqnarray}
These relations ensure the ``path-independence'' \cite{Teit}, i.e., the independence of the 4-geometry arising after integrating the equations of motion (\ref{eq:heq}) of the choice of  functions  $N(x^k,t), N^i(x^k,t)$ for fixed starting and end points of the evolution. Relations  (\ref{eq:nonalg}) do not define a Lie algebra in general, since they involve the dynamic variable $\gamma^{ki}$. However, for transformations of spatial coordinates (with $N=M=0$) we have the Lie-algebraic relation 
\be
\left\{H(N^i),
H(M^j)\right\}=H\left({[N,M]}^k\right), \label{eq:3diff} \ee
where 
\be
{[N,M]}^k=N^iM^k_{\ ,i}-M^iN^k_{\ ,i}.
\ee
It is natural to expect that (\ref{eq:nonalg}) and (\ref{eq:3diff}) are independent of the choice of variables and are preserved under changes of the variables. One such change of variables is considered in the next Section. 
In the case where the hypersurface has a boundary (including an infinitely remote one), all of the above requires a more thorough analysis. The Hamiltonian may differ from (\ref{eq:ham}) by surface integrals over the boundary, as is the case, for example, in  \cite{RT,Sol85}. We consider this case later.

\section{Ashtekar's transformation}

Instead of the metric tensor $\gamma_{ij}$ we introduce the triad $E^a_i$ in such a way that $\gamma_{ij}=E^a_iE^a_j$, $a=1,2,3$. The inverse matrices to the triad are denoted by $E^i_a$, hence, $E^a_iE^j_a=\delta^j_i$, and $E^a_iE^i_b=\delta^a_b$. Since $\gamma^{kj}\gamma_{ji}=
\gamma^{kj}E^a_jE^a_i=\delta^k_i$, the inverse matrix can be obtained by raising the index with the help of $\gamma^{kj}$, $E^k_a=E^{ka}=\gamma^{kj}E_j^a$. The position of the inner index $a$ is irrelevant. It is also not difficult to verify that \be
\gamma^{ij}=E^{ia}E^{ja},\qquad \gamma={\rm det}|E^a_iE^a_j|={\left(
{\rm det}|E^a_i|\right)}^2=E^2.
\ee
Let us introduce the momenta $\pi^i_a$ conjugate to the triad. They satisfy the equations 
\be
\left\{E^a_i(x),\pi^j_b(y)\right\}=\delta^j_i\delta^a_b\delta(x,y),
\ee and can be easily related to the momenta $\pi^{ij}$ by means of 
\be
\pi^{ij}=\frac{1}{4}\left(\pi^i_aE^j_a+\pi^j_aE^i_a\right).
\ee
It now turns out that part of the Poisson brackets for the ADM variables has been modified: 
\be
\left\{\gamma_{ij}(x),\pi^{kl}(y)\right\}=\delta^{kl}_{ij}\delta(x,y),\qquad
\left\{\gamma_{ij}(x),\gamma_{kl}(y)\right\}=0,
\ee while 
\be
\left\{\pi^{ij}(x),\pi^{kl}(y)\right\}=\frac{1}{4}\left(\gamma^{ik}{\cal M}^{jl}+
\gamma^{il}{\cal M}^{jk}+\gamma^{jk}{\cal M}^{il}+\gamma^{jl}{\cal M}^{ik}\right)\delta(x,y),
\ee
where 
\be
{\cal M}^{ij}=\frac{1}{4}\left(E^{ia}\pi^j_a-E^{ja}\pi^i_a\right)={\cal M}^{[ij]}.
\ee
To preserve the correspondence between Poisson structures, one has to impose three constraints ${\cal M}^{ij}=0$, which also ensures the conservation of the number of degrees of freedom (a symmetric tensor $\gamma_{ij}(x)$ is defined by six numbers at each point, while the triad matrix $E^a_i(x)$ contains nine independent components). The constraints can be represented equivalently in the form 
\be
J^{ab}\equiv
J^{[ab]}=0,\quad {\rm where} \quad J^{ab}={\cal M}^{ij}E_i^aE_j^b.
\ee
 The constraints are in involution, 
 \be
 \left\{{\cal M}^{ij}(x), {\cal
 M}^{kl}(y)\right\}=\frac{1}{4}\left(\gamma^{ik}{\cal M}^{jl}
 -\gamma^{il}{\cal M}^{jk}-\gamma^{jk}{\cal M}^{il}+
 \gamma^{jl}{\cal M}^{ik}\right)\delta(x,y).
\ee
Clearly, the choice of $\left(E^a_i,\pi^i_a\right)$ as the canonical variables is not unique. In view of the transition to the Ashtekar variables that we make below, it is more convenient to use the variables $\left(\tilde E^{ia},K_i^a\right)$ defined by 
\be
\tilde E^{ia}=EE^{ia},\quad K^a_i=K_{ij}E^{ja}+E^{-1}E_{ib}J^{ab},
\label{eq:EK}
\ee
Then, 
\be
\left\{\tilde E^{ia}(x),K^b_j(y)\right\}=\frac{1}{2}\delta^i_j\delta^{ab}
\delta(x,y),
\ee
\be
\left\{\tilde E^{ia}(x),E^{jb}(y)\right\}=0,\qquad
\left\{K^a_i(x),K^b_j(y)\right\}=0.
\ee
In  \cite{Ash} Ashtekar proposed a beautiful transformation that allowed one to represent the density of the gravitational Hamiltonian as a fourth-order polynomial in canonical variables. The presentation in this paper follows  \cite{HNS}. The Ashtekar transformation is analogous to the canonical transformations in classical mechanics, 
 \be
 q^A\rightarrow q^A,\quad p_A\rightarrow p_A+\frac{\partial F(q)}{\partial
 q_A}.
\ee
In the present case, the generating function $F(q)$  is replaced by the functional 
\be
F=F\left(\tilde E^{ia}\right)=\int\limits_{\Omega}\tilde E^{ia}\Gamma^a_id^3x,
\ee
where 
\be
\Gamma^a_i=\frac{1}{2}\epsilon^{abc}E_{jc}E^{jb}_{\ |i}.
\ee
and the partial derivative with respect to the coordinate is replaced by the Euler-Lagrange variational derivative 
\be
\frac{\delta F}{\delta\tilde E^{ia}}=\Gamma^a_i.
\ee
Ignoring the surface terms, this transformation can be viewed as a canonical one. Ashtekar also introduced a complex parametrization in which the new variables are represented as 
 \be
 A^a_i=iK^a_i+\Gamma^a_i,
\ee
In this parameterization, we have 
\be
\left\{\tilde
E^{ia}(x),A^b_j(y)\right\}=\frac{i}{2}\delta^i_j\delta^{ab}\delta(x,y),
\ee
\be
\left\{\tilde
E^{ia}(x),\tilde E^{jb}(y)\right\}=0, \qquad
\left\{A^a_i(x),A^b_j(y)\right\}=0.
\ee

Up to the surface terms that we consider in the subsequent Sections, changing the variables in the Hamiltonian leads to the expression 
\be
H=\int\limits_{\Omega}\left({\cal N} \epsilon^{abc}\tilde E^{ia}\tilde
E^{jb}F^c_{ij}+ N^i2i\tilde E^{ja}F^a_{ij}+\hat\xi^a2{\cal D}_i\tilde E^{ia}
\right)d^3x.\label{eq:HamAsh}
\ee
Here
\be
{\cal D}_i\tilde E^{ia}\equiv i\epsilon^{abc}J^{bc}\equiv
i\epsilon^{abc}E_{ib}E_{jc}{\cal M}^{ij},
\ee
 and the new covariant derivative ${\cal D}_i$ is defined by 
\be
{\cal D}_i\lambda^{ka}=\lambda^{ka}_{\ |i}+\epsilon^{abc}A^b_i\lambda^{kc}.
\ee
 The curvature of the connection $A^a_i$ can be found from 
 \be
 \left({\cal D}_i{\cal D}_j-{\cal D}_j{\cal D}_i\right)\lambda^a=
 \epsilon^{abc}F^b_{ij}\lambda^c,
\ee
hence 
\be
F^a_{ij}=\partial_iA^a_j-\partial_jA^a_i+\epsilon^{abc}A^b_iA^c_j.
\ee
Let us note that in this paper, as well as in  \cite{Sol92}, the constraints and the Hamiltonian differ from those given in \cite{Ash} by a factor of $2$, while the Poisson bracket differs by the factor of $1/2$, so, the equations of motion 
are identical. We prefer to use the current notations since the Lagrangian multiplier $N^i$ then coincides with the ADM-formalism multiplier, while ${\cal N}=E^{-1}N$. 
In order to make a comparison with the ADM-formalism  (\ref{eq:algconstr}), we give one more set of algebraic relations for the generators entering (\ref{eq:HamAsh}): 
\begin{eqnarray}
\{H({\cal N},N^i,\hat\xi^a),H({\cal M},M^j,\hat\eta^b)\}&=&
H({\cal L},L^k,\hat\lambda^c),\\
{\cal L}&=&N^k{\cal M}_{,k}-{\cal M}N^k_{\ ,k},\\
L^k&=&\tilde E^{ka}\tilde E^{ja}\left({\cal N}{\cal M}_{,j}-{\cal M}{\cal N}_
{,j}\right)+N^jM^k_{\ ,j}-M^jN^k_{\ ,j},\\
\hat\lambda^c&=&-i\epsilon^{cab}\hat\xi^a\hat\eta^b+\nonumber\\
&+&\frac{i}{2}F^c_{jk}(N^jM^k-M^jN^k)
+\epsilon^{cab}F^a_{jk}\tilde E^{kb}N^j{\cal M}.\\
\end{eqnarray}
In what follows, we are interested in the situation where the surface terms play an essential role. In this case, it is shown that the Ashtekar transformation differs from a canonical one, namely 
\be
\left\{A^a_i(x),A^b_j(y)\right\}\ne 0.
\ee

\section{Surface terms and  new formula for  Poisson brackets}

Let us recall that, from the geometrical point of view, the Poisson bracket arises from the following construction: 
\be
\{F,G\}=\d G\inprod\d F\inprod\Psi,\label{eq:geom}
\ee
where $\d$ is a differential (1-form), $\Psi$  is the Poisson bivector (whose Schouten-Nienhuis bracket with itself vanishes), and $\inprod$ denotes the inner product operation (in the present case, the inner multiplication of 1-forms with a bivector or a 1-vector). In field theory, one usually assumes\footnote{In this paper, we try to avoid the multi-index notations $J=(j_1,\dots,j_n)$,
$|J|=j_1+\dots+j_n$, and $D_J=\frac{\partial^{|J|}}{\partial^{j_1}
x^1\dots\partial^{j_n} x^n}$,  which are widely used in  \cite{Sol93, Sol95}. We give the formulae in the multi-index notation only in the present Section along with the standard notations in order to illustrate how much space can be saved.} 
that 
\begin{itemize}
\item
$F, G$  are local functionals, 
\be
F=\int\limits_{\Omega}f\left(\phi_A(x), D_i\phi_A,
D_iD_j\phi_A,\dots \right)d^nx=
\int\limits_{\Omega}f\left(\phi^{(J)}_A(x)\right)d^nx,
\ee
\item 
their differential is determined by the variational Euler-Lagrange derivative\footnote{By $D_i$ we denote the total partial derivative
$D_i={\frac{\partial}{\partial x^i}}+\phi_A^{(J+i)}{\frac{\partial}{\partial\phi_A^{(J)}}}$ and $D_J=D_1^{i_1}\ldots D_n^{i_n}$.}, 
\begin{eqnarray}
\d F&=&\int\limits_{\Omega}\frac{\delta F}{\delta\phi_A}\delta\phi_Ad^nx,\\
\frac{\delta F}{\delta\phi_A}&=&\frac{\partial f}{\partial\phi_A}-
D_i\frac{\partial f}{\partial D_i\phi_A}+D_iD_j\frac{\partial f}
{\partial D_iD_j\phi_A}+\dots=(-1)^{|J|}D_J\frac{\partial f}
{\partial\phi_A^{(J)}},\label{eq:E-L}
\end{eqnarray}
\item 
the coefficients $\hat I_{AB}$ of the Poisson bivector 
\be
\Psi=\frac{1}{2}\int\limits_{\Omega}\frac{\delta}{\delta\phi_A}\wedge\hat I_{AB}
\frac{\delta}{\delta\phi_B}d^nx,
\ee
are determined from the fundamental brackets among the fields, which are also local, i.e., 
\be
\{\phi_A(x),\phi_B(y)\}=\hat I_{AB}\delta(x,y),
\ee
\item 
the inner product of a 1-form and, for example, a 1-vector (which can always be brought to the canonical form through integrating by parts) reads 
\be
\alpha\inprod\psi=\left(\int\limits_{\Omega}\alpha_A\delta\phi_Ad^nx\right)\inprod
\left(\int\limits_{\Omega}\psi_B\frac{\delta}{\delta\phi_B}d^nx\right)=
\int\limits_{\Omega}\alpha_A\psi_Ad^nx.\label{eq:int_prod}
\ee
\end{itemize}
The Poisson bracket that results from (\ref{eq:geom}) is given by 
\be
\{F,G\}=\int\limits_{\Omega}\frac{\delta F}{\delta\phi_A}\hat I_{AB}
\frac{\delta G}{\delta\phi_B}d^nx,\label{eq:brack}
\ee 
where, in the simplest case of the canonical variables $\phi_A=(q_{\alpha},p_{\alpha})$ the operator $\hat I_{AB}$ is the totally antisymmetric matrix 
\be
%\begin{displaymath}
\hat I=
\left( \begin{array}{cc}
0 & \delta^{\alpha}_{\beta}\\
-\delta^{\alpha}_{\beta} & 0
\end{array}\right),
%\end{displaymath}
\ee
while, in the general case, $\hat I_{AB}$ is an antisymmetric differential operator of an arbitrary finite order with the coefficients depending on the the fields $\phi_A$ and their derivatives (which are also of a finite order),
\be
\hat I_{AB}=-{\hat I_{BA}}^{\ast},
\ee
where 
\begin{eqnarray}
\hat I_{AB}=I^{(0)}_{AB}+I^{(i)}_{AB}D_i+I^{(ij)}_{AB}D_iD_j+\dots=
I^K_{AB}D_K, \\
{\hat
I_{AB}}^{\ast}=I^{(0)}_{BA}-D_i\circ
I^{(i)}_{BA}+D_iD_j\circ I^{(ij)}_{BA}-\dots =(-1)^{|K|}D_K\circ I^K_{BA}.
\end{eqnarray} 
The above formula for the Poisson bracket  (\ref{eq:brack})  can be given a rigorous derivation in field theory (in the framework of the so-called formal variational calculus \cite{Olv, Dorf, Dick}) where ``good'' boundary conditions are imposed, i.e., when any integral of a total divergence vanishes. 

In physics, however, and in the Hamiltonian formalism for gravity, in particular, one encounters problems where this condition is not satisfied. For example, Hamiltonian (\ref{eq:ham}) in asymptotically flat space-time has to be supplemented by surface integrals of a special form  \cite{RT, Sol85}. It turns out, however, that one can still use the standard formula for the Poisson bracket if one is restricted to the class of ``differentiable'' Hamiltonians, whose variations do not involve surface integrals. In that case, one can preserve the definition of the differential in terms of the Euler-Lagrange derivative. Article \cite{BH}  was devoted to demonstrating that under the asymptotic boundary conditions at the spatial infinity adopted in \cite{RT, Sol85} the Poisson bracket does not map outside the class of ``differentiable'' functionals. By a direct check, one can see that the Jacobi identity holds under the conditions of  \cite{RT, Sol85}. 

At the same time, there remain several obscure points in the cited papers. For example, when one evaluates the Poisson brackets of the ADM-formalism generators in accordance with (\ref{eq:brack})  under the boundary conditions from \cite{RT,Sol85}, one obtains similar generators with the necessary surface terms included, and therefore, the algebra closes similarly to Eqs. (\ref{eq:alg}) and (\ref{eq:nonalg}). On the other hand, evaluating the bracket according to such formulae as
\be
\left\{H(N), H(M)\right\}=\int\limits_{\Omega}\int\limits_{\Omega}N(x)M(y)\{{\cal H}(x),
{\cal H}(y)\}d^3xd^3y\label{eq:localbrack}
\ee
with the help of the algebra of constraints  (\ref{eq:algconstr}), does not allow one to obtain the necessary surface integrals. The reason is that integrals of the type 
\be
\int\limits_{\Omega}\int\limits_{\Omega}\xi(x)\eta(y)
{\left(\frac{\partial}{\partial x}\right)}^m
{\left(\frac{\partial}{\partial y}\right)}^n\delta(x,y),
\ee
which emerge in the intermediate calculations, are only defined when all of the surface integrals vanish, which is not the case under the boundary conditions taken in \cite{RT, Sol85}. 

In the mid-eighties, publications appeared \cite{LMMR,FT85,APP} that went beyond the above class of ``differentiable'' functionals. Under certain boundary conditions, the functionals considered therein could have variations involving surface components. An inevitable consequence was that the standard formula for the Poisson brackets was invalidated, since it violated the Jacobi identity. Other formulas have been proposed that differ from the conventional one by boundary terms as well. 

Recently, we were able to show  \cite{Sol95} that it is possible to consider nontrivial boundary problems when the formal variational calculus  \cite{Olv, Dorf, Dick}  is generalized to include total divergences. In this approach, the Poisson bracket determined by Eq.  (\ref{eq:geom}) in the general geometrical setting satisfies the standard requirements of antisymmetry, the Jacobi identity, and closedness of its domain of definition ${\cal A}$, i.e., $F,G\in {\cal A}\rightarrow \{F,G\}\in {\cal A}$ , where ${\cal A}$ is the set of all local functionals, irrespective of the choice of boundary conditions. 

A new formula for the Poisson bracket emerges as a result of generalizing all of its three ``components'': the differential, the Poisson bivector, and the pairing. This is necessary if one wishes to preserve all of the surface terms in the expression for the functionals,  $m$-forms, and  $m$-vectors. The differential of a local functional is now given by its {\it total} variation, which does not imply dropping the boundary contributions, 
\be
\d F=\int\limits_{\Omega}f'_A\delta\phi_Ad^nx,
\label{eq:new_differential}
\ee 
where we introduced the Fr\'echet derivative 
\be
f'_A=\frac{\partial f}{\partial\phi_A}+\frac{\partial f}{\partial\phi_{A,i}}
D_i+\frac{\partial f}{\partial\phi_{A,ij}}D_iD_j+\dots=\frac{\partial f}
{\partial\phi^{(J)}_A}D_J.
\ee
Instead of partial derivatives, it is often convenient to use symmetrized covariant derivatives, which allows us to write 
\be
f'_A=\frac{\partial f}{\partial\phi_A}+\frac{\partial
f}{\partial(\nabla_i\phi_A)}\nabla_i+\frac{\partial
f}{\partial(\nabla_{(i}\nabla_{j)}\phi_A)}\nabla_{(i}\nabla_{j)}+\dots
\ee
Another useful rewriting is achieved when using the higher Euler operators \cite{Olv}, 
\begin{eqnarray}
\d F&=&\int\limits_{\Omega}\left(E^0_A(f)\delta\phi_A+
D_i\bigl(E^{1,i}_A(f)\delta\phi_A\bigr)+
D_iD_j\bigl(E^{2,ij}_A(f)\delta\phi_A\bigr)+\dots\right)d^nx=\nonumber\\
&=&\int\limits_{\Omega}D_J\left(E^J_A(f)\delta\phi_A\right)d^nx,\qquad
E^J_A(f)=(-1)^{|J|+|K|}{K\choose J}D_{K-J}\frac{\partial f}{\partial
\phi_A^{(K)}},
\end{eqnarray} 
where the zero-order operator $E^0_A$ is the standard Euler-Lagrange variational derivative  (\ref{eq:E-L}). A third way of writing these relations involves a formal trick that reduces the integral over a finite domain $\Omega$ to an integral over the entire infinite space $\Rn$, 
\be
F=\int\limits_{\Omega}fd^nx=\int\limits_{\Rn}\theta_{\Omega}fd^nx,
\ee
where 
\be
\theta_{\Omega}(x)=\cases{
1 &{\rm if} $x\in\Omega$\cr
0 &{\rm otherwise}\cr}.
\ee
In what follows, we omit the subscript $\Omega$ from the notation for this function and the symbol $\Rn$ from the notation for the integral over the entire space. We now obtain 
\be
\d F=\int E^0_A(\theta f)\delta\phi_Ad^nx=\int\frac{\delta F}
{\delta\phi_A}\delta\phi_Ad^nx,
\ee
and it also turns out that 
\be
E^0_A(\theta f)=\frac{\delta F}{\delta\phi_A}=
\theta E^0_A(f)-\theta_{,i} E^{1,i}_A(f)+\theta_{,ij}
E^{2,ij}_A(f)+\dots=(-1)^{|J|}D_J\theta E^J_A(f).\label{eq:fullvar}
\ee 
The new {\it total} variational derivative --- for which we use the same notation as the one normally used for the Euler-Lagrange derivative --- contains information not only about the integrand $f$, but, also, due to the presence of the $\theta$ function, about the integration domain $\Omega$. 

The second step consists in revising the  definition of the bivector. This, too, amounts to taking into account the characteristic function $\theta$ of the domain $\Omega$. Then, we have to modify the definition of the conjugate operator used when ``carrying over'' the derivatives from one of the arguments of the $\delta$-function to another: 
\be
\hat I_{AB}(x)\delta(x,y)={\hat I_{BA}}^{\ast}(y)\delta(x,y).
\label{eq:new_adjoint}
\ee
For example,
\be
\theta (x)D_x\delta(x,y)=-\theta (y)D_y\delta(x,y)
-\theta '\delta(x,y),
\ee
which means that whenever we have 
 \be
 \hat I=\theta D,
\ee
then 
\be
{\hat I}^{\ast}=-\theta D-\theta '=-D\circ\theta .
\ee
Taking such terms into account allows one, in particular, to avoid ambiguities and to obtain consistent answers when using formulas of type (\ref{eq:localbrack}) and  (\ref{eq:brack}). As an antisymmetric operator (in the sense of the new conjugation operation) one should, clearly, use the expression $(1/2)(\hat I-\hat I^\ast )$. 
Finally, the pairing $\inprod$ of 1-forms and bivectors (in the general case, of $m$-vectors) or, conversely, of 1-vectors and 2-forms (in general, $m$-forms) is also defined anew, using the trace of two differential operators. If 
\begin{eqnarray} 
\hat A&=&a_JD_J\equiv
a+a_iD_i+a_{ij}D_iD_j+\dots,\\ 
\hat B&=&b_KD_K\equiv
b+b_kD_k+b_{kl}D_kD_l+\dots,
\end{eqnarray}
then 
\begin{eqnarray} 
{\rm Tr}
\left(\hat A\hat B\right)&=&ab+a_iD_ib+D_kab_k
+D_ka_iD_ib_k+D_kD_lab_{kl}+D_kD_la_iD_ib_{kl}+\nonumber\\
&+&a_{ij}D_iD_jb+D_ka_{ij}D_iD_jb_k+
D_kD_la_{ij}D_iD_jb_{kl}+\dots=D_Ka_JD_Jb_K.
\end{eqnarray} 
In this case, the generalization of Eq. (\ref{eq:int_prod}) reads as 
\be
\alpha\inprod\psi=\left(\int\limits_{\Omega}\hat\alpha_A\delta\phi_Ad^nx\right)
\inprod
\left(\int\limits_{\Omega}\hat\psi_B\frac{\delta}{\delta\phi_B}d^nx\right)=
\int\limits_{\Omega}{\rm Tr}\left(\hat\alpha_A\hat\psi_A
\right)d^nx.\label{eq:new_int_prod}
\ee 

As shown in  \cite{Sol95}, the above three steps --- generalizing the definitions of the differential (\ref{eq:new_differential}), the conjugate operator  (\ref{eq:new_adjoint}), and the pairing  (\ref{eq:new_int_prod})  --- considered together with the most general geometrical definition (\ref{eq:geom}), result in the new formula for the Poisson bracket. In the calculations that follow, we use the representation through the Fr\'echet derivatives 
\be
\{F,G\}=\int{\rm Tr}\left(f'_A\hat I_{AB}g'_B\right)d^nx,\label{eq:newbrack}
\ee
where $\hat I_{AB}$ is an antisymmetric operator and the $\theta $-functions are pulled outside the trace sign in the  operator $I_{AB}$, i.e.. if 
\be
\hat I_{AB}=
\theta \hat I^{\langle  0\rangle }_{AB}+
\theta_{,i} \hat I^{\langle  i\rangle }_{AB}+
\theta_{,ij} \hat I^{\langle  ij\rangle }_{AB}+\dots
=D_J\theta \hat I^{\langle  J\rangle }_{AB},
\ee
then 
\begin{eqnarray}
\{F,G\}&=&\int\limits_{\Omega}{\rm Tr}\left(f'_A\hat I_{AB}^{\langle  0
\rangle }g'_B\right)d^nx-
\int\limits_{\Omega}D_i{\rm Tr}\left(f'_A\hat I_{AB}^{\langle  i
\rangle }g'_B\right)d^nx+
\int\limits_{\Omega}D_iD_j{\rm Tr}\left(f'_A\hat I_{AB}^{\langle  ij
\rangle }g'_B\right)-
\dots=\nonumber\\
&=&(-1)^{|J|}\int\limits_{\Omega}D_J{\rm Tr}\left(f'_A\hat I_{AB}^{\langle  J
\rangle }g'_B\right)d^nx.
\end{eqnarray}
 The result can be written, using the higher Euler derivatives, as 
\be
\{F,G\}=(-1)^{|L|}
\int\limits_{\Omega}D_{J+K+L}\left(E^J_A(f)\hat I^{\langle  L\rangle
} E^K_B(g)\right)d^nx,
\ee
 or as the double integral involving total variational derivatives (\ref{eq:fullvar}), 
\be \{F,G\}=\int\int\frac{\delta
F}{\delta\phi_A(x)}\frac{\delta G}{\delta \phi_B(y)}\{\phi_A(x),\phi_B(y)\}
d^nxd^ny,
\ee
which should be evaluated according to the rules explained in \cite{Sol93}. 

\section{Surface terms in  ADM formalism} 

As our first example of the evaluation of Poisson brackets according to Eq. (\ref{eq:newbrack}), we consider the bracket of the functionals known as the generators of spatial diffeomorphisms in asymptotically flat space-time, 
\be
\left\{H(N^i),H(M^j)\right\}=\int\limits_{\Omega}{\rm Tr}\left(h'_{\gamma_{ij}}
(N^i)h'_{\pi^{ij}}(M^j)-h'_{\gamma_{ij}}
(M^j)h'_{\pi^{ij}}(N^i)\right)d^3x,
\ee
where
\be
H(N^i)=\int\limits_{\Omega}N^i{\cal H}_id^3x+
\oint\limits_{\partial\Omega}2\pi^j_iN^idS_j
\equiv\int\pi^{ij}(N_{i|j}+N_{j|i})d^3x.
\ee
The variation of such a functional reads 
 \be
\delta H=\int\left((N_{i|j}+N_{j|i})\delta\pi^{ij}+2\pi^{ij}\delta
(\gamma_{ik}(N^k_{\ ,j}+\Gamma^k_{jm}N^m))\right)d^3x.
\ee 
Using the known formula 
\be
\delta\Gamma^k_{jm}=\frac{1}{2}\gamma^{kn}\left(\delta\gamma_{nj|m}+
\delta\gamma_{nm|j}-\delta\gamma_{jm|n}\right),
\ee
 we arrive at 
\be
\delta H(N^i)=\int\limits_{\Omega}\left((N_{i|j}+N_{j|i})\delta\pi^{ij}+
(\pi^{ik}N^j_{\ |k}+\pi^{kj}N^i_{\ |k})\delta\gamma_{ij}+N^k\pi^{ij}(\delta
\gamma_{ij})_{|k}\right)d^3x,
\ee
which shows that the Fr\'echet derivatives with respect to the canonical variables are equal to 
 \begin{eqnarray}
 h'_{\pi^{ij}}(N^i)&=&N_{i|j}+N_{j|i},\\
 h'_{\gamma_{ij}}(N^i)&=&\pi^{ik}N^j_{\ |k}+\pi^{kj}N^i_{\ |k}
 +N^k\pi^{ij}\nabla_k,
\end{eqnarray}
where $\nabla_k$ and the vertical line denote the same covariant derivative (the one compatible with the metric tensor $\gamma_{ij}$). Therefore, the Fr\'echet derivative of the generator with respect to the momenta is a function, while the derivative with respect to the metric tensor is a differential operator. Unlike in the calculations performed when evaluating the Euler-Lagrange derivative, there is no need to integrate by parts in our case. Let us evaluate the trace 
\be
{\rm Tr}\left(h'_{\gamma_{ij}}(N^i)h'_{\pi^{ij}}(M^j)\right)=
\left(\pi^{ik}N^j_{\ |k}+\pi^{kj}N^i_{\ |k}+N^k\pi^{ij}\nabla_k\right)
(M_{i|j}+M_{j|i}).
\ee
The terms that are symmetric with respect to $N, M$ do not contribute to the Poisson bracket. After renaming the indices, we obtain 
\be
\left\{H(N^i),H(M^j)\right\}=\int\limits_{\Omega}2\pi^{ij}\left(
(N^k_{\ |j}M_{i|k}-
M^k_{\ |j}N_{i|k})+(N^kM_{i|jk}-M^kN_{i|jk})\right)d^3x.
\ee
We now change the order of the second covariant derivatives using the relation 
\be
M_{i|jk}=M_{i|kj}+R_{mijk}M^m,
\ee
Then the contribution of 
\be
2\pi^{ij}R_{mijk}\left(N^kM^m-M^kN^m\right)
\ee
vanishes by virtue of symmetry properties of the Riemann tensor and, thus, 
\be
\left\{H(N^i),H(M^j)\right\}=\int\limits_{\Omega}2\pi^{ij}{\left(N^kM_{i|k}-
M^kN_{i|k}\right)}_{|j}d^3x=H([N,M]^k).\label{eq:adm}
\ee
It can be seen that the generators  $H(N^i)$ realise a representation of the algebra of diffeomorphisms of a three-dimensional hypersurface. Note that we have not specified any special boundary conditions and, thus, these generators would not be ``differentiable'' functionals in the standard approach. From our point of view, this simply means that the standard formula for the Poisson bracket cannot be used in the general case. If one attempts to formally evaluate the same bracket using the conventional formula  (\ref{eq:brack}) the result would differ from ours by the surface integral 
\be
\Delta \left\{H(N^i),H(M^j)\right\}=
-\oint\limits_{\partial\Omega}\pi^{ij}\left(N^k(M_{i|j}+M_{j|i})-
M^k(N_{i|j}+N_{j|i})\right)dS_k.
\ee
 
Naturally, the Jacobi identity for the bracket (4.4) would not be satisfied, in general: 
\[
\left\{\{H(N^i),H(M^j)\},H(L^k)\right\}+
\left\{\{H(L^i),H(N^j)\},H(M^k)\right\}+
\left\{\{H(M^i),H(L^j)\},H(N^k)\right\}\ne 0.
\]
The condition ensuring the validity of the Jacobi identity for the conventional Poisson bracket consists in requiring that  $N^i, M^j, L^k$  be Killing vectors of the metric tensor  $\gamma_{ij}$  at the boundary  $\partial\Omega$ or  be tangent to the boundary. In the first case, obviously, the new and the old formulae produce the same result, even though the functionals  $H(N^i)$, $H(M^j)$  remain ``nondifferentiable'': 
\be
\delta H=\int\limits_{\Omega}\left(E^0_{\gamma_{ij}}(h)\delta\gamma_{ij}+
E^0_{\pi^{ij}}(h)\delta\pi^{ij}\right)d^3x+\oint\limits_{\partial\Omega}
N^k\pi^{ij}\delta\gamma_{ij}dS_k.
\ee

\section{Surface terms in  Ashtekar's formalism} 

It is natural to expect that going over from ADM to the Ashtekar variables should not affect the spatial diffeomorphism algebra (\ref{eq:adm}) in any considerable way. However, there are at least two subtleties of this transformation which we would like to discuss in a greater detail. 

First, as has already been noted, the transition to the triad preserves the original Poisson brackets only on the constraint surface  ${\cal M}^{ij}=0$ and, therefore, some extra calculations are required in order to derive the complete (off-shell) structure of the Poisson-bracket algebra. 

Second, as has also been mentioned above, the Ashtekar transformation is canonical only up to surface terms and it is important to understand the role played in the algebra by the noncanonical contribution. 

As long as we are talking about applications of the new formula for the Poisson bracket, we assume that it is justified to give technical details about the calculations. The general proof of the invariance of that formula under changing dependent variables (field redefinitions) has not been published yet and, therefore, an explicit demonstration of this invariance in a concrete example would not be redundant. Thus, we start with checking how relation (\ref{eq:3diff}) changes under going over to the triad. Since 
 \be
 \left\{\pi^{ij}(x),\pi^{kl}(y)\right\}={\cal C}^{ijkl}_{mn}
 {\cal M}^{mn}\delta(x,y),\label{eq:deform}
\ee
where
\be
{\cal C}^{ijkl}_{mn}=\frac{1}{4}\left(\gamma^{ik}\delta^{jl}_{mn}+
\gamma^{il}\delta^{jk}_{mn}+\gamma^{jk}\delta^{il}_{mn}+
\gamma^{jl}\delta^{ik}_{mn}\right),
\ee
the bracket found above receives an additional contribution that vanishes on the constraint surface  ${\cal M}^{mn}=0$
 \begin{eqnarray}
 \Delta\left\{H(N^i),H(M^j)\right\}&=&\int\limits_{\Omega}{\rm Tr}\left(
 h'_{\pi^{ij}}(N^i){\cal C}^{ijkl}_{mn}{\cal M}^{mn}h'_{\pi^{kl}}(M^j)
 \right)d^3x =\nonumber\\
 &=&\int\limits_{\Omega}(N_{i|j}+N_{j|i})
 {\cal C}^{ijkl}_{mn}{\cal M}^{mn}(M_{k|l}+M_{l|k})d^3x
 =\nonumber\\
 &=&\int\limits_{\Omega}(N^k_{\ |m}+N_m^{\ |k})(M_{k|n}+M_{n|k})
 {\cal M}^{mn}d^3x.\label{eq:triads}
\end{eqnarray}
This additional contribution may also vanish outside the constraint surface if at least one of the vector fields  $N^i(x), \ M^j(x)$ satisfies the Killing equation, i.e., preserves the metric tensor  $\gamma_{ij}(x)$  in the domain $\Omega$. 

In order to evaluate the bracket directly in the new variables, for instance in  variables $\left(E^a_i,\pi^i_a\right)$, it is not always necessary to use the explicit expressions for the generators in terms of these variables. It is often sufficient to express the variations of the old fields in terms of the new ones: 
\be
\delta\gamma_{ij}=E^a_i\delta E^a_j+E^a_j\delta E^a_i,
\ee
\be
\delta\pi^{ij}=\frac{1}{4}\left(E^{ia}\delta\pi^{ja}+\pi^{ja}\delta E^{ia}
+E^{ja}\delta\pi^{ia}+\pi^{ia}\delta E^{ja}\right).
\ee
Then 
\begin{eqnarray}
\delta H&=&\int\limits_{\Omega}\left(h'_{\gamma_{ij}}\delta\gamma_{ij}+
h'_{\pi_{ij}}\delta\pi^{ij}\right)d^3x=
\int\limits_{\Omega}\left(\tilde h'_{E_{ia}}\delta E_{ia}+
\tilde h'_{\pi^{ia}}\delta\pi^{ia}\right)d^3x=\nonumber\\
&=&\int\limits_{\Omega}\left(h'_{\gamma_{kj}}(\gamma_{kj})'_{E_{ia}}\delta
E_{ia}+h'_{\pi_{kj}}(\pi^{kj})'_{E_{ia}}\delta E_{ia}+
h'_{\pi_{kj}}(\pi^{kj})'_{\pi^{ia}}\delta\pi^{ia}\right)d^3x,
\end{eqnarray}
whence we find 
\begin{eqnarray}
\tilde h'_{\pi^{ia}}(N^i)&=&h'_{\pi_{ij}}(N^i)\frac{1}{2}E^{ja}=\frac{1}{2}
(N_{i|j}+N_{j|i})E^{ja},\\
\tilde h'_{E_{ia}}(N^i)
&=&h'_{\gamma_{ij}}(N^i)2E_{ja}+h'_{\pi^{kl}}(N^i)\left(
-\frac{1}{2}\right)\pi^{kb}E^{ib}E^{la}=\nonumber\\
&=&\left(\pi^{ik}N^l_{\ |k}+\pi^{kl}N^i_{\ |k}+N^k\pi^{il}\nabla_k\circ
\right)2E_{la}
-\frac{1}{2}(N_{k|l}+N_{l|k})\pi^{kb}E^{ib}E^{la}.
\end{eqnarray}
and 
\begin{eqnarray}
{\rm Tr}\left(\tilde h'_{E_{ia}}(N^i)\tilde h'_{\pi^{ia}}(M^j)\right)&=&
%\left(\pi^{ik}N^l_{\ |k}+\pi^{kl}N^i_{\ |k}+N^k\pi^{il}\nabla_k
%\right)2E_{la}\frac{1}{2}(M_{i|j}+M_{j|i})E^{ja}-\nonumber\\
%&-&\frac{1}{2}(N_{k|l}+N_{l|k})\pi^{kb}E^{ib}E^{la}
%\frac{1}{2}(M_{i|j}+M_{j|i})E^{ja}=\nonumber\\
\left(\pi^{ik}N^l_{\ |k}+\pi^{kl}N^i_{\ |k}+N^k\pi^{il}\nabla_k
\right)(M_{i|l}+M_{l|i})-\nonumber\\
&-&\frac{1}{4}\pi^{kb}E^{ib}(N_k^{\
|j}+N^j_{\ |k})(M_{i|j}+M_{j|i}).
\end{eqnarray} 
In this way, we have 
\begin{eqnarray}
\left\{H(N^i),H(M^j)\right\}&=&\int\limits_{\Omega}{\rm Tr}\left(
\tilde h'_{E_{ia}}(N^i)
\tilde h'_{\pi^{ia}}
(M^j)-
\tilde h'_{E_{ia}}(M^i)\tilde h'_{\pi^{ia}}(N^j)\right)d^3x=\nonumber\\
&=&H([N,M]^k)
+\int\limits_{\Omega}(N^k_{\ |m}+N_m^{\ |k})(M_{k|n}+M_{n|k}){\cal
M}^{mn}d^3x,
\end{eqnarray}
and, as was to be expected, the result does not change under this change of variables as compared with the ADM bracket (\ref{eq:triads}) deformed according to (\ref{eq:deform}), (\ref{eq:adm}). 

Similar considerations apply to going over to $\left(\tilde E^{ia},K^a_i\right)$. In this case, we use Eqs. (\ref{eq:EK}) and another relation 
\be
\pi^{ia}=2E\left(E^{ia}E^{jb}-E^{ib}E^{ja}\right)K^b_j,
\ee
which is easy to verify. We thus obtain 
\begin{eqnarray}
\delta E_{ia}&=&\frac{1}{2E}(E_{ia}E_{jb}-2E_{ja}E_{ib})\delta\tilde E^{jb},
\nonumber\\
\delta\pi^{ia}&=&2E(E^{ia}E^{jb}-E^{ib}E^{ja})\delta K^b_j+
\biggl(
2(K^b_jE^{ia}+K^c_kE^{kc}\delta^i_j\delta^{ab}-K^c_jE^{ic}\delta^{ab}-
\nonumber\\
&-&K^b_kE^{ka}\delta^i_j)+K^c_kE_{jb}(E^{ic}E^{ka}-E^{kc}E^{ia})
\biggr)\delta\tilde E^{jb}.
\end{eqnarray}
Calculations similar to those performed above also confirm, in this case, that the result is unchanged when one chooses new variables. 

Let us note that, up to this point, we have beep considering only algebraic transformations, i.e., those free of field derivatives. When going over to the Ashtekar variables 
\be
\tilde E^{ia}\rightarrow
\tilde E^{ia},\quad K^a_i\rightarrow A^a_i=iK^a_i+ \Gamma^a_i,
\ee
this is no longer the case, because 
\be
\Gamma^a_i=\frac{1}{2}\epsilon^{abc}\tilde E_{jc}\tilde E^{jb}_{\ |i}.
\ee
where $\tilde E_{jc}$ is the matrix inverse to $\tilde E^{ib}$. Expressing the variations of the old variables through the new ones we obtain 
\be
\delta K^a_i=-i\delta A^a_i+i{\left(\Gamma^a_i\right)}'_{\tilde E^{jb}}
\delta\tilde E^{jb},
\ee
For an arbitrary functional, therefore, we have 
\be
\delta F=\int\limits_{\Omega}\left(f'_{\tilde E^{ia}}\delta\tilde E^{ia}+
f'_{K^a_i}\delta K^a_i\right)d^3x=\int\limits_{\Omega}\left(
\tilde f'_{\tilde E^{ia}}\delta\tilde E^{ia}+
\tilde f'_{A^a_i}\delta A^a_i\right)d^3x,
\ee
where 
\be
\tilde f'_{\tilde E^{ia}}=f'_{\tilde E^{ia}}+f'_{K^b_j}{\left(\Gamma^b_j
\right)}'_{\tilde E^{ia}},
\ee
\be
\tilde f'_{A^a_i}=-if'_{K^a_i}.
\ee
If these transformations were canonical (up to the factor of $i$), the Poisson bracket would be given in the new variables by 
 \begin{eqnarray}
 \{F,G\}&=&\frac{i}{2}\int\limits_{\Omega}{\rm Tr}
 \left(\tilde f'_{\tilde E^{ia}}
 \tilde g'_{A^a_i}-\tilde f'_{A^a_i}\tilde g'_{\tilde E^{ia}}\right)d^3x=
 \frac{1}{2}\int\limits_{\Omega}{\rm Tr}\left(f'_{\tilde E^{ia}}
 g'_{K^a_i}-f'_{K^a_i}g'_{\tilde E^{ia}}\right)d^3x+\nonumber\\
 &+&
 \frac{1}{2}\int\limits_{\Omega}{\rm Tr}\left(f'_{K^b_j}{(\Gamma^b_j)}'_{\tilde E^{ia}}
 g'_{K^a_i}-f'_{K^a_i}g'_{K^b_j}{(\Gamma^b_j)}'_{\tilde E^{ia}}\right)d^3x.
\end{eqnarray}
 However, the invariance would be violated, 
\be
\Delta_1\{F,G\}=\int\limits_{\Omega}
{\rm Tr}\left(f'_{K^a_i}\hat C_{aibj}g'_{K^b_j}
\right)d^3x,
\ee
where 
\be
\hat C_{aibj}=\frac{1}{2}\left({(\Gamma^a_i)}'_{\tilde E^{jb}}-
{\left[{(\Gamma^b_j)}'_{\tilde E^{ia}}\right]}^{\ast}\right).
\ee
We can see that the hypothesis stating that the Ashtekar variables are canonical contradicts the invariance of the Poisson bracket. However, we showed  \cite{Sol92} --- still making use of the standard formula (\ref{eq:brack}) --- that 
\be
\left\{A^a_i(x),A^b_j(y)\right\}=i\hat C_{aibj}(x)\delta(x,y)\ne 0,
\ee
 precisely, 
 \begin{eqnarray}
 \left\{A^a_i(x),A^b_j(y)\right\}&=&i\left\{\Gamma^a_i(x),K^b_j(y)\right\}+
 i\left\{K^a_i(x),\Gamma^b_j(y)\right\}=
 \frac{i}{2}\biggl({\bigl(\Gamma^a_i(x)
 \bigr)}'_{\tilde E^{jb}}-\nonumber\\
 &-&{\bigl(\Gamma^b_j(y)
 \bigr)}'_{\tilde E^{ia}}\biggr)
 \delta(x,y)=\frac{i}{2}\left({\bigl(\Gamma^a_i(x)\bigr)}'_{\tilde
 E^{jb}}- {\biggl[{\bigl(\Gamma^b_j(x)\bigr)}'_{\tilde
 E^{ja}}\biggr]}^{\ast}\right)\delta(x,y).
\end{eqnarray}
If we ignore the surface terms, the Fr\'echet derivative of the Euler-Lagrange derivative is a symmetric operator \cite{Olv}, hence, we have a purely surface contribution in this case. Since the Ashtekar variables are noncanonical, a second correction arises in this way, 
\be
\Delta_2\{F,G\}=\int\limits_{\Omega}{\rm Tr}\left(\tilde f'_{A^a_i}\hat C_{aibj}
\tilde g'_{A^b_j}\right)d^3x=-\int\limits_{\Omega}{\rm Tr}\left(f'_{K^a_i}\hat
C_{aibj}g'_{K^b_j}\right)d^3x,
\ee
that compensates the first one. The new formula for the Poisson bracket remains invariant under the replacement $\left(\tilde E^{ia}, K^a_i\right)\rightarrow
\left(\tilde E^{ia}, A^a_i\right)$, which pertains to the surface contributions. 

The invariance of the bracket allows us to use different variables for calculating the algebra of the generators. Explicit calculations that make use of the generators in the Ashtekar formalism are much more involved than those considered above. For comparison, let us present the explicit expression in the Ashtekar variables of the $H(N^i)$ generator; the shortest derivation of this generator is given in  \cite{Thi}: 
\begin{eqnarray}
H(N^i)&=&2i\int\limits_{\Omega}N^k\tilde E^{ia}F^a_{ki}d^3x-
2i\oint\limits_{\partial\Omega}\bigl(A^a_i-\Gamma^a_i\bigr)\bigl(
\tilde E^{ia}N^k-\tilde E^{ka}N^i\bigr)dS_k-\nonumber\\
&-&2i\int\limits_{\Omega}\left(N^iA^c_i-\frac{1}{2}\epsilon^{abc}\tilde E_{ib}
\bigl(\tilde E^{ia}_{\ |k}N^k-\tilde E^{ka}N^i_{\ |k}
\bigr)\right){\cal D}_j\tilde
E^{jc}d^3x.
\end{eqnarray}
As an example of calculations performed directly in the Ashtekar variables, consider the bracket of the generators that implement complex rotations of the triad. The variation of the generator reads 
\be
\delta H(\hat\xi^a)=\delta\int\limits_{\Omega}
\hat\xi^a2{\cal D}_i\tilde E^{ia}d^3x=
\int\limits_{\Omega}\left(h'_{\tilde E^{jc}}(\hat\xi^a)\delta\tilde E^{jc}+
h'_{A^c_j}(\hat\xi^a)\delta A^c_j\right)d^3x,
\ee
where 
\begin{eqnarray}
h'_{\tilde E^{jc}}(\hat\xi^a)&=&2\hat\xi^c\partial_j+2\hat\xi^a\epsilon^{abc}A^b_j,\\
h'_{A^c_j}(\hat\xi^a)&=&-2\hat\xi^a\epsilon^{abc}\tilde E^{jb}.
\end{eqnarray}
The Poisson bracket is given by 
\begin{eqnarray}
\{H(\hat\xi^a),H(\hat\eta^b)\}&=&\frac{i}{2}\int\limits_{\Omega}{\rm Tr}
\left(h'_{\tilde E^{jc}}(\hat\xi^a)h_{A^c_j}(\hat\eta^b)-
h'_{A^c_j}(\hat\xi^a)h'_{\tilde E^{jc}}(\hat\eta^b)\right)d^3x+\nonumber\\
&+& i\int\limits_{\Omega}{\rm Tr}
\left(h'_{A^c_j}(\hat\xi^a)\hat C_{cjdk}h'_{A^d_k}(\hat\eta^b)\right)d^3x.
\end{eqnarray}
Here, one needs the explicit form of the noncanonical correction
\be
\hat C_{aibj}=\theta_{,k}C^k_{aibj}=\theta_{,k}\frac{i}{4E}\left(
\epsilon^{acb}{\delta^k}_jE_{ic}-\epsilon^{bca}{\delta^k}_iE_{jc}-
\epsilon^{acd}E_{ib}E_{jc}E^{kd}+\epsilon^{bcd}E_{ja}E_{ic}E^{kd}\right).
\ee
As a result, we obtain the following relations: 
\begin{eqnarray}
\{H(N^i),H(M^j)\}&=&H\left({[N,M]}^k\right)+H(\hat\lambda^a),
\label{eq:1}\\
\{H(N^i),H(\hat\xi^a)\}&=&0,\label{eq:2}\\
\{H(\hat\xi^a),H(\hat\eta^b)\}&=&H\left(-i{(\hat\xi\times\hat\eta)}^c\right)
\label{eq:alggauge},
\end{eqnarray}
with 
\be
{[N,M]}^k=N^iM^k_{\ ,i}-M^iN^k_{\ ,i}, \qquad {(\hat\xi\times\hat\eta)}^c=
\epsilon^{cab}\hat\xi^a\hat\eta^b,
\ee
\be
\hat\lambda^a=\epsilon^{abc}E^{ib}E^{jc}\left(N^k_{\ |i}+N_i^{\ |k}\right)
\left(M_{k|j}+M_{j|k}\right).
\ee
where the surface terms are fixed, while the boundary conditions remain completely free. 

The last term in (\ref{eq:1}) was missed in a similar formula, (5.3) of  \cite{Thi}, which, however, was not essential for the final result. The origin of this inaccuracy consists, of course, in the fact that going over from the canonical brackets for the ADM variables to the deformed brackets  (\ref{eq:deform}) is not a change of variables and preserves the Poisson brackets only on the constraint surface ${\cal M}^{ij}=0$. 

\section{Conclusions} 

We have shown how the new definition of the field theory Poisson brackets allows  one to analyse Poisson bracket algebras for a  class of functionals which is broader than  the standard one. We allow arbitrary local functionals rather than  ``differentiable'' functionals only (i.e., those whose variation does not contain boundary contributions). Moreover, it turns out that one can find generators, which, in general, are different from the constraints by surface integrals, in such a way that their algebra closes. The expressions found for the generators cannot be called new, because they appear, for example, in the analysis of the asymptotically flat space-time  \cite{Thi}. What is really new, however, is the statement regarding their applicability in a much more general context, irrespective to the choice of boundary conditions. In another paper  \cite{Sol96} we discussed the criterion for a local functional to be the admissible Hamiltonian, i.e., the criterion ensuring that this functional generates the standard equations of motion (rather than equations of motion ``in the weak sense''). In relation to the present paper, this, more restrictive, requirement would not change the Poisson algebra we found in (\ref{eq:1}), (\ref{eq:2}),  and (\ref{eq:alggauge}); rather, it would lead to the requirement that the displacement vectors of the spatial coordinates be tangent to the boundary. 

The present paper is limited to the discussion of the algebra of generators of the spatial diffeomorphisms and the triad rotations, i.e., transformations acting within the same hypersurface. It is not yet clear what would happen when the hypersurface starts moving in the normal direction, or whether it is possible to organize the corresponding generators into the algebra similar to the one well-known for spaces without boundaries or for asymptotically flat spaces. Answering these questions is equivalent to finding out whether there exists a surface integral that would allow one to close this more general algebra involving the generator  $H({\cal N})$ for any boundary conditions.

\end{document}